\documentclass[pre,amsmath,amssymb,twocolumn,superscriptaddress]{revtex4}
\usepackage{amsmath,amssymb,graphicx}
\usepackage{dcolumn}

\begin{document}


\title{Thermal transport in long-range interacting Fermi-Pasta-Ulam chains}

\author{Jianjin Wang}
\affiliation{Department of Physics, Jiangxi Science and Technology
Normal University, Nanchang 330013, Jiangxi and Department of
Physics, Fuzhou University, Fuzhou 350108, Fujian, China}
\author{Sergey V. Dmitriev}
\affiliation{Institute for Metal Superplasticity Problems of RAS,
Khalturin Street 39, 450001 Ufa, Russia}
\affiliation{Institute of
Mathematics with Computing Centre, the Ufa Federal Research Centre
of RAS, Chernyshevsky Street 112, 450008 Ufa, Russia}

\author{Daxing Xiong}
\email{xmuxdx@163.com} \email{phyxiongdx@fzu.edu.cn}
\affiliation{School of Science, Jimei University, Xiamen 361021 and
Department of Physics, Fuzhou University, Fuzhou 350108, Fujian,
China}

\begin{abstract}
Studies of thermal transport in long-range (LR)interacting systems
are currently particularly challenging. The main difficulties lie in
the choice of boundary conditions and the definition of heat current
when driving systems in an out-of-equilibrium state by the usual
thermal reservoirs. Here, by employing a reverse type of thermal
baths that can overcome such difficulties, we reveal the intrinsic
features of thermal transport underlying a LR interacting
Fermi-Pasta-Ulam chain. We find that under an appropriate range
value of LR exponent $\sigma =2$, while a \emph{nonballistic}
power-law length ($L$) divergence of thermal conductivity $\kappa$,
i.e., $\kappa \sim L^{\alpha}$ still persists, its scaling exponent
$\alpha \simeq 0.7$ can be much larger than the usual predictions in
short-range interacting systems. The underlying mechanism is related
to the system's new heat diffusion process, weaker nonintegrability
and peculiar dynamics of traveling discrete breathers. Our results
shed light on searching for low-dimensional materials supporting
higher thermal conductivity by involving appropriate LR
interactions.
\end{abstract}

\maketitle

\section{Introduction}
Thermal transport in one-dimensional (1D) systems has been a topic
of both theoretical and practical interest for several
decades~\cite{Book2,LepriReport,DharReport}. One of the main
achievements has been to realize that most 1D systems can display
anomalous behaviors. This anomaly means that the ``standard''
thermal transfer law, i.e., Fourier's law, $J=- \kappa \nabla T$,
which states that heat current $J$ is proportional to temperature
gradient $\nabla T$ with $\kappa$ the thermal conductivity being
constant for a bulk material, is not valid. Instead, $\kappa$ shows
a sublinear power-law divergence as increasing system size $L$,
i.e., $\kappa \sim L^{\alpha}$ ($0<\alpha<1$). This issue is in part
motivated by the recent carbon nanotubes (CNTs)
technology~\cite{CNTs-1,CNTs-2}, and also by the desire to
understand the microscopic origin of non-Fourier's thermal
transport. Additionally, a fascinating topic of efficiently
controlling heat has led to the emerging field of
phononics~\cite{Phononics-1,Phononics-2,Phononics-3,Phononics-4,Phononics-5}.
Nevertheless, despite these achievements and the practical
importance, our main theoretical understandings are based on the 1D
anharmonic chains with nearest-neighbor (NN) couplings, whose
interactions are short-range (SR). How thermal transport would
behave in long-range (LR) interacting systems is an open fundamental
question.

LR interactions are usually characterized by potential $V(r)$ decays
with the distance $r$ between two particles in a power law $V(r)
\propto r^{-\sigma}$, where $\sigma$ denotes the range of
interaction. Generally, the cases of $\sigma$ lower than the
system's spatial dimension (in 1D systems, $\sigma <1$) are called
as the LR-interacting systems~\cite{Book,Report2009}. Such
LR-interactions are ubiquitous in several physical systems, ranging
from self-gravitating to nanoscale, and to quantum
systems~\cite{Book,Report2009,Report2010,Report2014}. These LR
interacting systems can display peculiar features, such as ensemble
inequivalence~\cite{Report2009}, anomalous diffusion of
energy~\cite{Report2010} and lack of additivity~\cite{Report2014}.
The nonadditivity means that even in the thermodynamic limit, to
decompose an initial equilibrium system into two effectively
noninteracting subsystems is impossible. As a result, when one deals
with thermal transport by coupling the usual thermal reservoirs to
two ends of the system~\cite{Lepri1997}, the central bulk system and
the two thermalized ends are implicitly identified. Eventually, the
whole system may display properties that do not correspond to the
original one. In this respect one does not know how to choose the
boundary conditions and define the heat current~\cite{Iubini2018}.
This difficulty has limited our understanding of thermal transport
of a true LR interacting system~\cite{Iubini2018,
Avila2015,Olivares2016,Bagchi2017-1,Bagchi2017-2,Cintio2019,Saito2019}.

In this article we aim to reveal the intrinsic feature of thermal
transport underlying a LR interacting chain. Toward that end and
following~\cite{Iubini2018,
Avila2015,Olivares2016,Bagchi2017-1,Bagchi2017-2,Cintio2019,Saito2019},
we use the terminology of LR interactions in a wider sense
regardless of the range exponent $\sigma$. This means that we
regarding systems with all $\sigma$ values as the generalized LR
interacting systems. One might immediately recognize that for
$\sigma \geq 1$ the nonadditivity will not appear, however, we
stress that in studies of thermal transport, the difficulties of
choice of the boundary conditions and definition of heat current
still persist when one applies the usual thermal reservoirs. To
solve these difficulties, we here instead employ a ``reverse
nonequilibrium molecular dynamics (RNEMD)''
method~\cite{RNEMD,Xiong2012} to build the nonequilibrium stationary
state. This method is advantageous since it can get rid of the
problem of boundary effects. Besides, as shown below, the definition
of heat current is natural. With this advantage we are able to
reveal that a length-divergence exponent of $\alpha \simeq 0.7$,
which is much larger than the original theoretical predictions
($\alpha \simeq 0.2$-$0.5$,
see~\cite{Book2,LepriReport,DharReport,Beijeren2012,Spohn2014}), can
be achieved in a theoretical model of the LR interacting
Fermi-Pasta-Ulam (FPU) chain under an appropriate range value
$\sigma=2$. This provides the theoretical possibility of producing a
higher $\alpha$ ($\alpha \simeq
0.6$-$0.8$)~\cite{CNTs-1,CNTs-3,LeiWang2002} by including suitable
LR interactions. This also suggests the need to extend the study of
thermal transport in systems beyond SR
interactions~\cite{Saito2019}. In principle, this implies new
mechanisms for transport.

Before proceeding, we would also like to emphasize that the present
technology already allows us to fabricate materials with LR
interactions. The Coulomb crystals~\cite{Examples-1}, Ising
pyrochlore magnets~\cite{Examples-2, Examples-3}, and permalloy
nanomagnets~\cite{Examples-4} are some of the notable examples.
Highly efficient thermal rectification has recently been achieved in
systems involving LR interactions~\cite{Rect-1,Rect-2}.

\section{Models}
We consider a FPU type LR interacting chain of $N$ particles with
Born-von Karman periodic boundary condition~\cite{Born¨CvonKarman}
whose dynamics is governed by the Hamiltonian~\cite{qs}:
\begin{equation}
H=\sum_{i}^N \left[\frac{p_{i}^{2}}{2}+ \frac{1}{2} \sum_{j\neq i}^N \frac{(x_{j}-x_i)^2}{(r_{ij})^{\sigma'}} + \frac{1}{4} \sum_{j\neq i}^N \frac{(x_j-x_i)^4}{(r_{ij})^\sigma}  \right].
\label{HH}
\end{equation}
Here, both the particle's mass and lattice constant are set to be
unity and all the relevant quantities are dimensionless; $x_{i}$ and
$p_{i}$ are two canonically conjugated variables with $i$ the index
of the particle; $(r_{ij})^{\sigma' (\sigma)}$ represents the
interaction strength of the $i$th particle with its $|j-i|$th
neighbors with $\sigma'(\sigma)$ the range value of LR exponent.
$r_{ij}$ is the shortest distance between particles $i$ and $j$ on
this periodic chain, and we define
\begin{equation}
r_{ij}=\left\{
\begin{aligned}
0 & , & i=j, \\
|j-i| & , & 1 \leq |j-i| \leq N/2, \\
N-|j-i| & , & N/2+1 \leq |j-i| \leq N-1.
\end{aligned}
\right.
\end{equation}

We do not include the Kac scaling factors $\tilde{N'}
(\tilde{N})=\frac{1}{N} \sum_{i=1}^N \sum_{j \neq i}^N
(r_{ij})^{-\sigma' (-\sigma)}$. These factors were designed to
restore the system's extensivity as increasing system size, but it
does not help improve the system's nonadditivity~\cite{Report2010}.
It only constructs an ``artificial'' extensive system, but the cost
for thermal transport is that both the phonon's group velocity and
the strength of nonlinearity should depend on $N$, which is an
unwanted effect. In fact, if one looks at dynamical aspects, for a
correspondence between both treatments, the difference is that time
should be $\tilde{N}^{1/2}$ scaled (setting
$\tilde{N'}=\tilde{N}$)~\cite{Tamarit2000}.

We shall focus on the case of $\sigma=2$ ($\sigma'=\infty$) and also
present the result of $\sigma=8$ ($\sigma'=\infty$) for comparison.
This only involves the LR interactions in the quartic anharmonic
term~\cite{Bagchi2017-1}, which differs from the model
in~\cite{Iubini2018}, but we have confirmed that it does not violate
our general conclusion~\cite{Note}. $\sigma=8$ might correspond to
the original FPU model with NN interactions, although this
equivalence requires $\sigma \rightarrow \infty$. The case of
$\sigma=2$ is particularly interesting since in this case ballistic
transport ($\alpha \simeq 1$)~\cite{Bagchi2017-1,Iubini2018}, like
that observed in integrable systems~\cite{Lebowitz1967,Toda1979},
has been conjectured. As the anharmonicity will generally cause
nonintegrablity, one might attribute such observed integrable
dynamics to the strong finite-size effects suffered from using the
usual thermal reservoirs~\cite{Lepri1997} and including the Kac
scaling factors. Indeed, it has been pointed out, due to this, that
to gain convincing results for $\sigma=2$ is extremely
hard~\cite{Iubini2018}. But anyway, as shown in the following, this
implies new physics. Another point for $\sigma=2$ of interest is
that the system under zero temperature can support tail-free
traveling discrete breathers (DBs)~\cite{Doi2016}. Then one might
ask, what will happen to these moving excitations under
finite-temperature systems and how would they affect transport?
Addressing this would also invoke the broad interest of the DB
field.

\section{Method}
The RNEMD method produces a temperature gradient in a reverse way.
Unlike the traditional approach~\cite{Lepri1997} to directly induce
$\nabla T$, it imposes the heat current by frequently exchanging
particle kinetic energy (or momentum). While the nonequilibrium
stationary state is reached, $\nabla T$ will be established. This
``reversion'' makes the RNEMD method an ideal candidate for studying
thermal transport in LR interacting systems. We also note that the
vasp~\cite{Vasp} and lammps~\cite{Lammps} codes modified to such
method have been widely used.

We implement the method in this way: First, since the Born-von
Karman periodic boundary condition is used, the chain forms like a
ring. We decompose the ring into $M=32$ equal slabs (each contains
$n = N/M$ particles). We give each slab a serial number $k$ and
label the cold one slab $1$ and accordingly, the hot one slab
$M/2+1$. This labeling allows us to interchange the momentum of the
hottest particle in slab $1$ with that of the coldest particle in
slab $M/2 + 1$ at a frequency $f_{exc}=0.1$. Such interchanges cause
a redistribution of kinetic energy of the system with an energy
difference: $\Delta E= \sum \frac{p_h^2-p_c^2}{2}$ during time $t$,
where the subscripts $h$ and $c$ refer to the hottest and coldest
particles whose momenta are exchanged, and the sum runs over all
exchange events in time $t$. As a consequence, the relaxation of
energy difference will drive two heat currents to flow from hot to
cold slabs along the two semi-ring sides bridging them (each side
has an effective length $L = N/2-n$). After the stationary state is
eventually reached, the long-time averaged current across each side
is defined by $\langle J \rangle=\lim_{t \rightarrow \infty}
\frac{\Delta E}{2t}$; accordingly the time averaged kinetic
temperature of slab $k$ is $\langle T_k \rangle=\frac{1}{n k_B}
\sum_{i=n(k-1)+1}^{nk} p_i^2$, where $k_B$ is the Boltzmann constant
(set to unity) and the sum runs over all $n$ particles in slab $k$.
The heat conductivity $\kappa$ can then be obtained by
$\kappa=-\langle J \rangle / \nabla T$ according to Fourier's law,
with $\nabla T$ being evaluated over the slabs between the cold and
hot ones.

We start calculations with several fully thermalized systems under
$T=0.5$. These systems are evolved by the velocity-Verlet
algorithm~\cite{Verlet} with a small time step $ 0.01$, that
guarantees energy conservation with a relative accuracy of
$O(10^{-5})$. We adopt a Fast Fourier Transform (FFT)~\cite{FFT}
algorithm to accelerate our computations. This helps avoid $O(N^2)$
operations in calculating forces at each time step in LR interacting
systems. With this, a transient time $10^6$ for the system to reach
the stationary state, is discarded, and another evolution time
$10^6$ is performed for the average.
\begin{figure}[!t]
\vskip-0.2cm
\includegraphics[width=8.8cm]{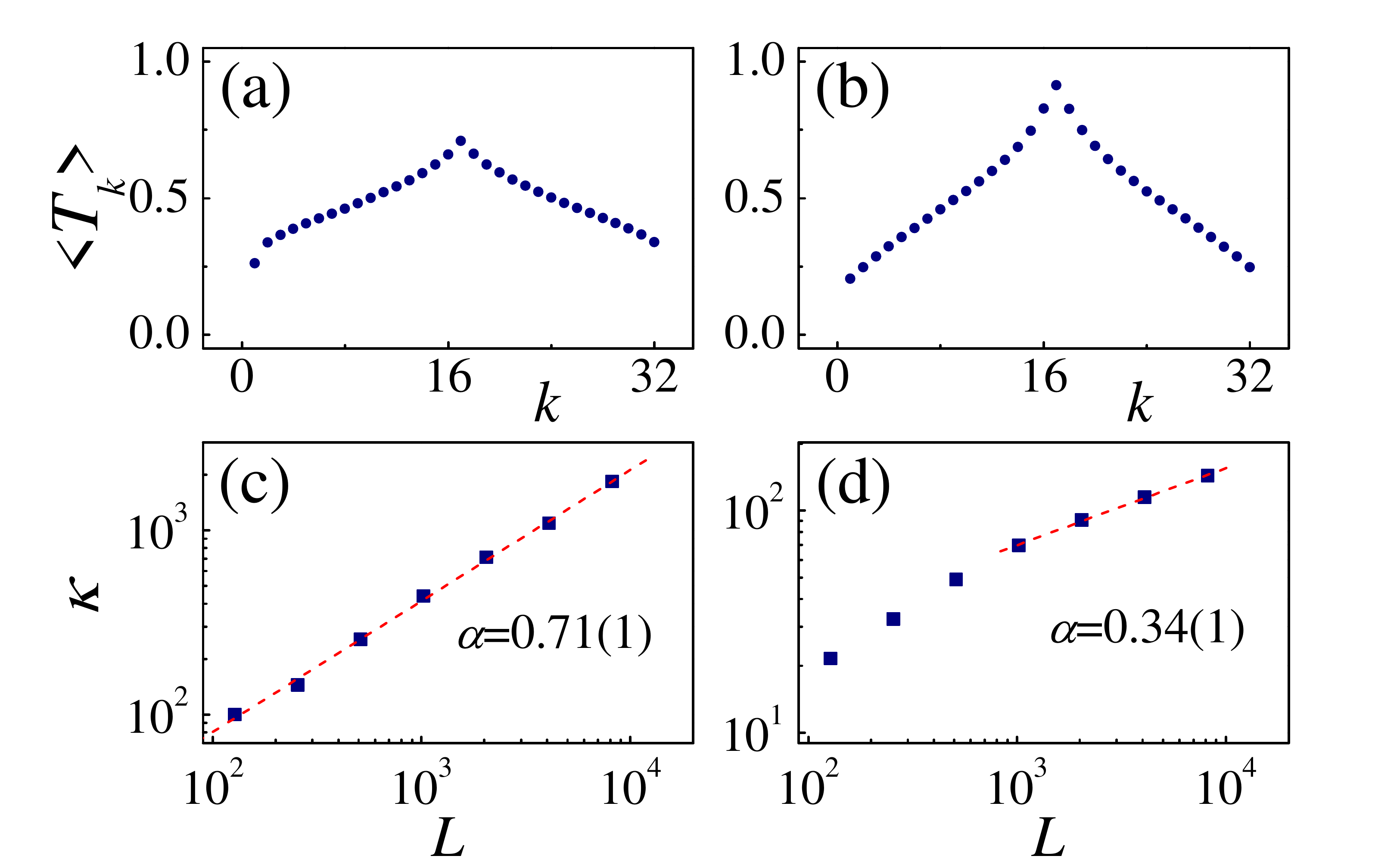}
\vskip-0.3cm
\caption{Temperature profile over slabs for $\sigma=2$ (a)
and $8$ (b) with $L = 7680$ ($N=16384$). $\kappa$ vs $L$ for $\sigma=2$ (c) and $8$ (d). The dashed lines denote $\kappa \sim L^\alpha$, suggesting $\alpha  = 0.71 \pm 0.01$ ($0.34 \pm 0.01$) for $\sigma=2$ (c) [$8$ (d)].} \label{fig:1}
\end{figure}

\section{Main results}
Figures~\ref{fig:1}(a,b) depict two typical temperature profiles
over slabs for $\sigma=2$ and $8$. In both cases a well-behaved
temperature gradient is identified. The difference is that $\nabla
T$ for $\sigma=2$ is noticeably smaller than that for $\sigma=8$.
Despite this difference, both results are obviously not the flat
temperature profiles found in integrable
systems~\cite{Lebowitz1967,Toda1979}. Therefore, here the
(quasi-)integrable dynamics is excluded.

Figures~\ref{fig:1}(c,d) show the result of $\kappa(L)$. As usual,
$\kappa \sim L^{\alpha}$ is observed. This is the case for
$\sigma=2$ for the entire considered range of $L$, while for
$\sigma=8$ the asymptotic behavior can only be achieved for large
$L$ as the crossover to the NN interaction model. Indeed, the best
fitting gives $\alpha=0.34 \pm 0.01$, which is close to the
prediction of $\alpha=\frac{1}{3}$ in FPU chains with NN
interaction~\cite{Narayan2002} and within the recent two predicted
universality classes~\cite{Beijeren2012,Spohn2014}. In contrast, for
$\sigma=2$, we obtain an enhanced $\kappa$ (at least one order of
magnitude compared to $\sigma=8$ for the same $L$) and a quite large
$\alpha$ ($\simeq 0.71$). We note that our estimation is convincing
as it is already at a very large scale ($N=16384$). In fact, such
calculations for only the last points of Figs.~\ref{fig:1}(c,d) take
one month of CPU time by an intel Xeon E5-2697v4 core even with the
FFT technique~\cite{FFT}. We emphasize that this $\alpha$, even
larger than usual, is clearly at variance with $\alpha=1$, again
supporting the system's nonintegrable dynamics. More-importantly,
this larger $\alpha$ indicates a higher thermal conductivity,
undoubtedly of potential applications.
\begin{figure}[!t]
\vskip-0.2cm
\includegraphics[width=8.8cm]{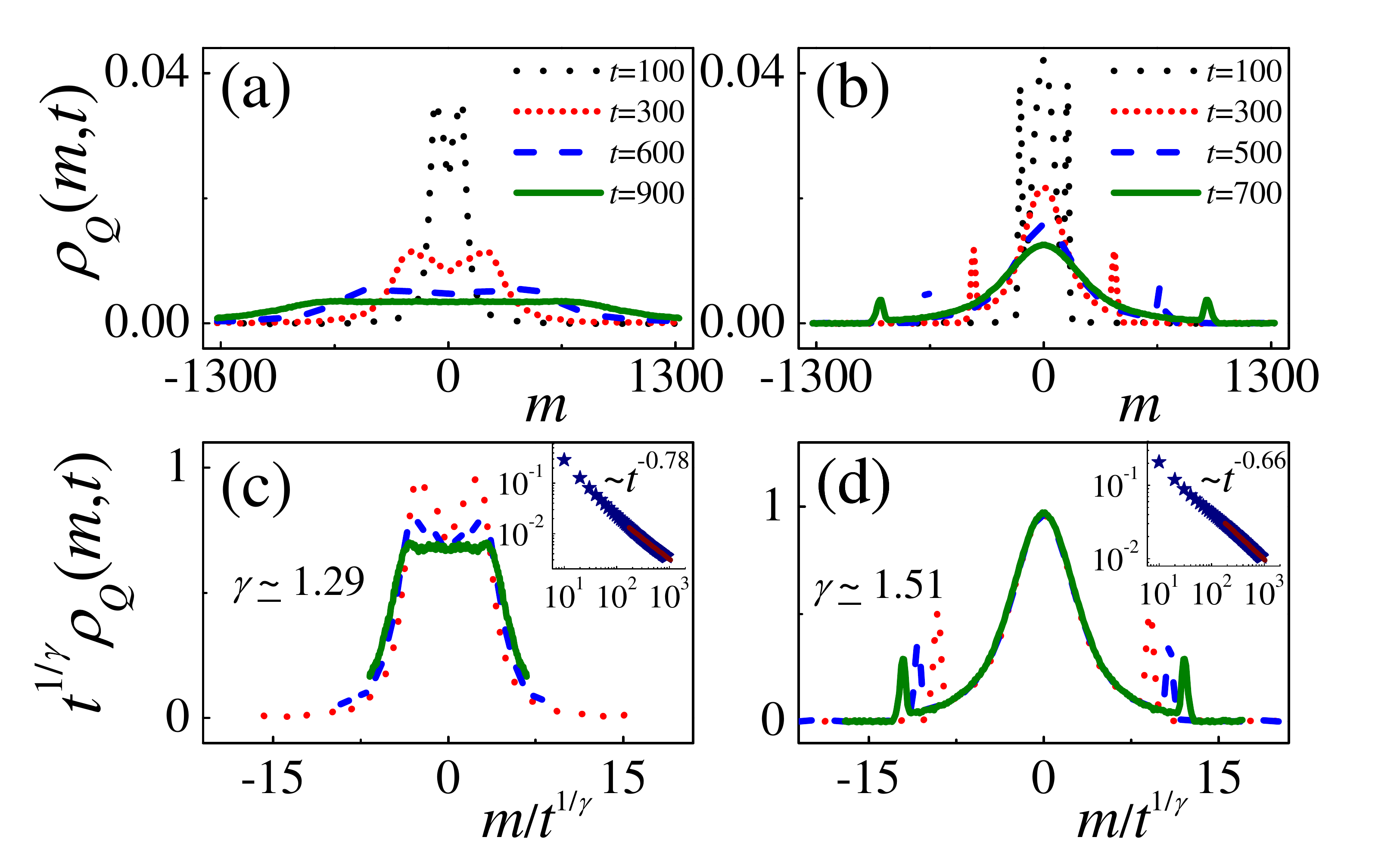}
\vskip-0.4cm
\caption{$\rho_Q(m,t)$ for several $t$ for $\sigma=2$ (a) and $8$ (b). (c) and (d) give the corresponding rescaled $\rho_Q(m,t)$  [as $t^{1/\gamma} \rho_Q(m/t^{1/\gamma},t)$]. The inset shows $\rho_Q(0,t)$ vs $t$ to derive the exponent $\gamma$.} \label{fig:2}
\end{figure}

\section{Underlying mechanisms}
\subsection{New heat diffusion process}
We explore this new exponent's underlying mechanisms from the
following three aspects. First, a new heat diffusion process is
revealed. This is exhibited in the propagation of heat fluctuations
following a new shaped density with a new scaling. To characterize
such process, we employ the equilibrium spatiotemporal correlation
function~\cite{Zhao2006,Zhao2013}
\begin{equation}
\rho_{Q}(m,t)=\frac{\langle \Delta Q_{l+m}(t) \Delta Q_{l}(0) \rangle}{\langle \Delta Q_{l}(0) \Delta Q_{l}(0) \rangle}
\end{equation}
of local thermal energy $Q_l(t)=E_l(t)-\frac{(\langle E \rangle +
\langle F\rangle) g_l(t)}{\langle g \rangle}$. Here, due to the
translational invariance, the correlation depends only on the
relative distance $m$; $\langle \cdot \rangle$ represents the
spatiotemporal average; $l$ labels a coarse-grained bin's number
similar to that adopted in the RNEMD method (each bin has $n=8$
particles). In the definition of $Q_l(t)$, $g_l(t)$ is the particle
number density, $E_l(t)=\sum_{k} E_k(t)$ with
$E_k=\frac{p_{k}^{2}}{2}+ \frac{1}{2} (x_{k+1}-x_k)^2+ \frac{1}{4}
\sum_{j \neq k}^N \frac{(x_j-x_k)^4}{|j-k|^\sigma}$ is the energy
density (note that here $\sigma'=\infty$ and the sum runs over all
particles within bin $l$), and $F_l(t)(\langle F\rangle \equiv 0)$
is the pressure density, respectively. To evaluate these densities,
one can calculate the number of particles $g_l(t)$, the total energy
$E_l(t)$ in the bin, and the pressure $F_l(t)$ exerted on the bin,
in each time $t$. Note that since the system is one dimensional the
pressure is equal to the force and can be computed from the gradient
of the potential.

Figures~\ref{fig:2}(a,b) depict $\rho_{Q}(m,t)$ for several $t$. The
calculations are performed under several equilibrium states of
$T=0.5$ and $N=4096$. For $\sigma=8$, as usual there are L\'{e}vy
walk-like profiles~\cite{Report2015} with a slowly relaxed central
peak together with two ballistically moving side peaks [see
Fig.~\ref{fig:2}(b)], like what is observed in the SR interacting
models. In contrast, for $\sigma=2$, $\rho_{Q}(m,t)$ shows a new
shape [see Fig.~\ref{fig:2}(a)]. Remarkably, the central peak for a
long time now turns to a platform, indicating a much faster
relaxation. We have verified that the short time's $\rho_{Q}(m,t)$
behaves similarly to that shown in ballistic
transport~\cite{Cintio2019}, however our long-time scaled dynamics
does support superdiffusive transport. Therefore, thanks to
excluding the Kac scaling factor, the real regime of transport is
revealed.

We figure out this new sort of transport by studying $\rho_Q(0,t)$
with $t$ (see the insets of Fig.~\ref{fig:2}). It shows a good
scaling $\rho_Q(0,t) \sim t^{-1/\gamma}$ with $\gamma \simeq
\frac{1}{0.78} \simeq 1.29$, in contrast to $\gamma \simeq
\frac{1}{0.66} \simeq 1.51$ for $\sigma=8$. With this, the rescaled
$\rho_{Q}(m,t)$ is plotted by
\begin{equation} \label{Scaling}
t^{1/\gamma} \rho(m,t) \simeq \rho(t^{-1/\gamma}m,t).
\end{equation}
This scaling formula is based on L\'{e}vy walk
theory~\cite{Report2015}. As shown, it is well satisfied for
$\sigma=8$ (especially the central parts). For $\sigma=2$ [see
Fig.~\ref{fig:2}(c)], to see an excellent collapse requires a much
longer time. However, this does not preclude the use of such a
formula. In fact, another relation
$\alpha=2-\gamma$~\cite{Report2015} based on the same theory
connecting $\gamma$ to $\alpha$ just gives an excellent estimation
$\alpha \simeq 2-1.29=0.71$, in agreement with our above thermal
conduction calculation.
\begin{figure}[!t]
\vskip-0.4cm
\includegraphics[width=7cm]{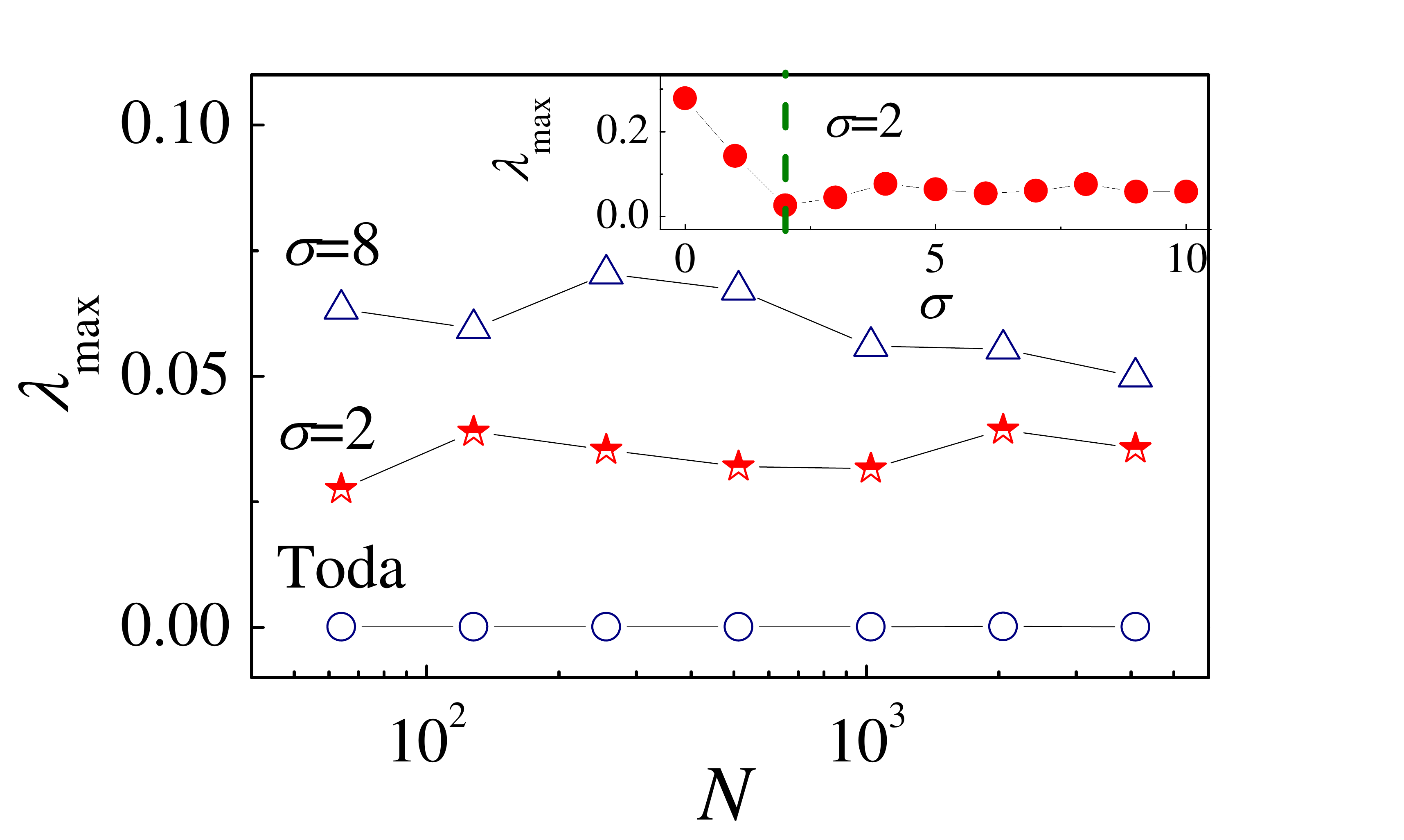}
\vskip-0.3cm
\caption{$\lambda_{\rm max}$ vs system size $N$ for Toda chain, $\sigma=2$, $8$ (from bottom to top). The inset shows $\lambda_{\rm max}$ vs $\sigma$.} \label{fig:3}
\end{figure}

\subsection{System's weaker nonintegrability}
Second, $\alpha \simeq 0.71$ seemingly relates the system's weaker
nonintegrability. The nonintegrability is featured by the maximal
Lyapunov exponent $\lambda_{\rm max}$ ($>0$) (see Fig.~\ref{fig:3}),
obtained from the standard Benettin-Galgani-Strelcyn
technique~\cite{Ly}. As a comparison, another completely integrable
Toda chain~\cite{Toda1979,Toda1} with $\lambda_{\rm max}=0$ is also
demonstrated. As shown, $\lambda_{\rm max}$ for $\sigma=2$ just lies
in between the results of $\sigma=8$ and the Toda chain, and this
seems unchanged with further increasing system size. It thus
indicates the system's weaker nonintegrability compared to the
counterpart SR interacting systems. Indeed, the nonmonotonic
variation of $\lambda_{\rm max}$ on $\sigma$ (see the inset) also
confirms this~\cite{Bagchi2017-1}, but the integrable dynamics are
certainly ruled out. Therefore, a weaker nonintegrability seems to
provide a mechanism to raise the divergent exponent.

\subsection{Peculiar dynamics of traveling DBs}
\begin{figure}[!t]
\vskip-0.2cm
\includegraphics[width=8.8cm]{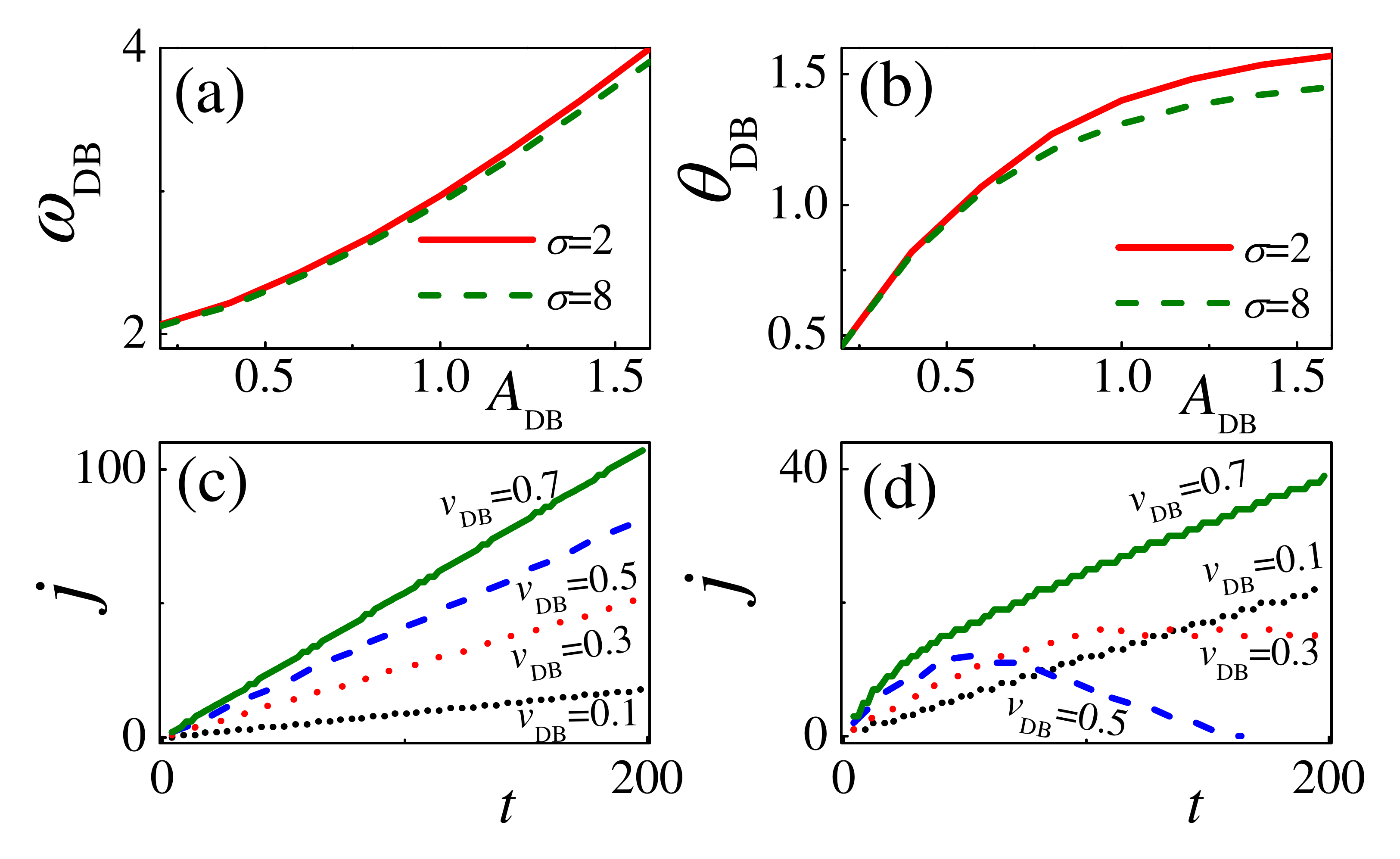}
\vskip-0.4cm
\caption{$\omega_{\rm DB}$ (a) and $\theta_{\rm DB}$ (b) vs $A_{\rm DB}$. DB's coordinate $j$ vs $t$ for $\sigma=2$ (c) and $8$ (d) for several $v_{\rm DB}$ for $A_{\rm DB}=1.6$.} \label{fig:4}
\end{figure}

Third, we conjecture that this weaker nonintegrability makes the
system support a new type of excitations--the tail-free traveling
DBs~\cite{Doi2016}--and it is these moving DBs together with their
relatively weak interactions that contribute to the higher
divergence. To verify this is interesting but usually greatly
challenging since it is hard to catch out these moving excitations
at equilibrium states due to their mobility. Viewing this we choose
to first present the dynamics of moving DBs in zero-temperature
systems (see Fig.~\ref{fig:4}). This is explored by using the
following ansatz~\cite{DBMethod}:
\begin{equation}
x_j(t)=\frac{(-1)^j A_{\rm DB} \cos[\omega_{\rm DB} t+ v_{\rm DB} (j-x_0)]}{\cosh[\theta_{\rm DB} (j-x_0-v_{\rm DB} t)]}. \label{DB}
\end{equation}
Here $A_{\rm DB}$ ($\omega_{\rm DB}$; $v_{\rm DB}$; $\theta_{\rm
DB}$) parametrizes DB's amplitude (frequency; velocity; inverse
width), and $x_0$ is the initial position of the DB. For a standing
DB with $v_{\rm DB}=0$, one can set $A_{\rm DB}$ and find
$\theta_{\rm DB}$ using a trial and error method~\cite{DBMethod},
which minimizes the oscillations of $A_{\rm DB}$. As soon as
$\theta_{\rm DB}$ has been obtained, we then calculate $\omega_{\rm
DB}$. This gives a general relation for both $\omega_{\rm DB}$ and
$\theta_{\rm DB}$ versus $A_{\rm DB}$ [see Figs.~\ref{fig:4}(a,b)].
The moving DB can then be excited by applying Eq.~\eqref{DB} as the
initial condition for the chosen $A_{\rm DB}$ and $v_{\rm DB}$. As
examples, we measure several DBs' real propagations under $A_{\rm
DB}=1.6$ for several $v_{\rm DB}$ [see Figs.~\ref{fig:4}(c,d)]. As
shown, DBs for $\sigma=2$ can move freely for all studied initial
$v_{\rm DB}$, but in contrast, the velocities of DBs for $\sigma=8$
decrease and they can stop. This clearly demonstrates the
distinction of DBs between $\sigma=2$ and $8$.
\begin{figure}[!t]
\includegraphics[width=9cm]{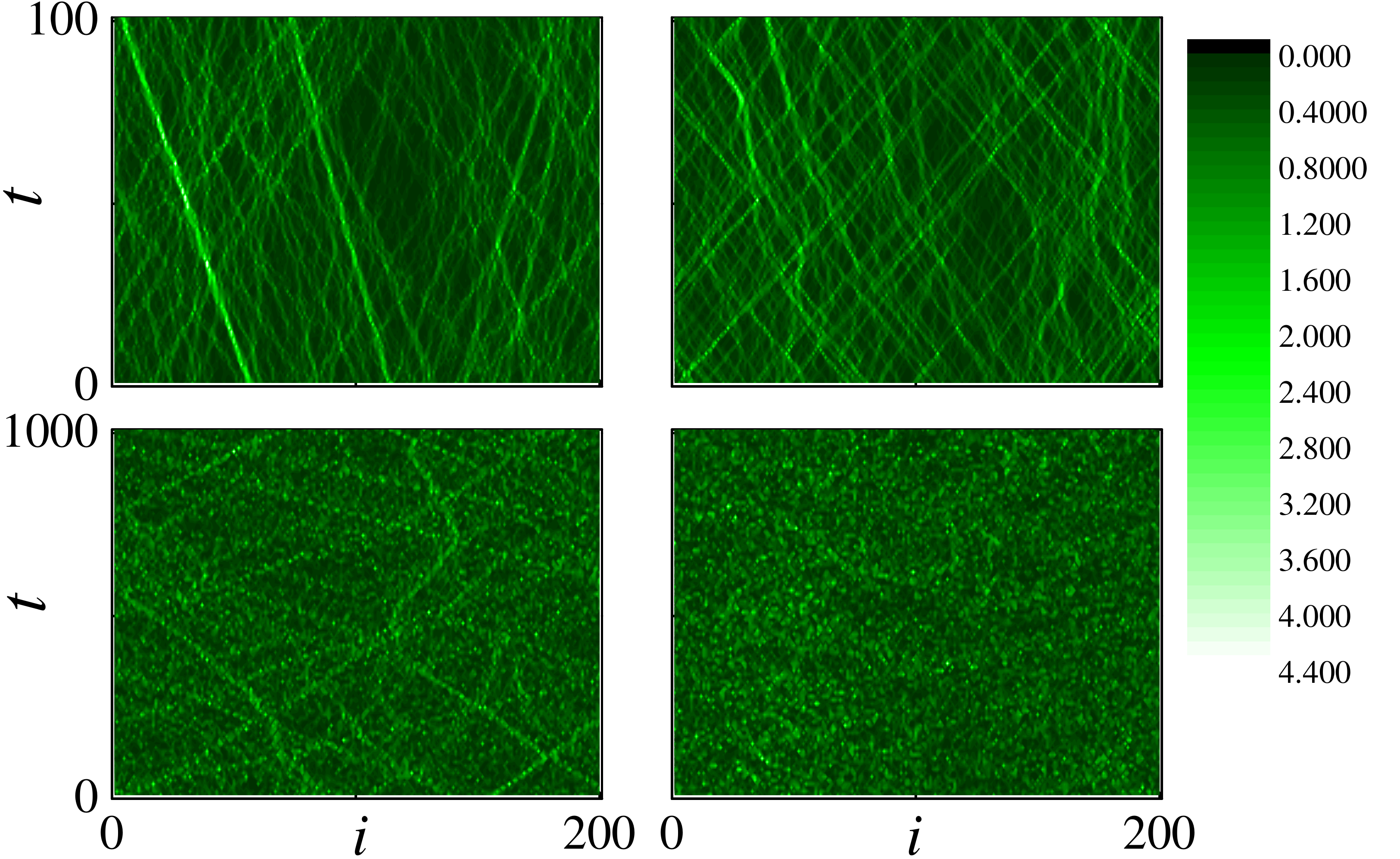}
\vskip-0.2cm \caption{Spatiotemporal evolution of energy densities
$E_i(t)$ at thermal equilibrium. Left (right) panels: $\sigma=2$
($8$). Upper (lower) panels: short ($t=100$) [long ($t=1000$)] time
with time step $\Delta t=1$ ($10$).} \label{fig:5}
\end{figure}

We secondly study the spatiotemporal evolutions of local energy
densities $E_i(t)$ under equilibrium states for two timescales
($t=100$ and $1000$) to visualize DBs' interactions in
finite-temperature systems (see Fig.~\ref{fig:5}). The evolutions
are obtained by considering a short chain with $N=200$. The chain is
first thermalized to $T=0.5$; then the thermal baths are removed and
the results are recorded and displayed by a suitable time step
[$\Delta t=1(10)$ for $t=100(1000)$]. As indicated, both the
$\sigma=2$ and $8$'s short timescale dynamics exhibit transport
similar to the ballistic regime. This explains the ballistic scaling
observed in a short time. In contrast, for a relatively long-time
scale, the ballistic transport for $\sigma=8$ disappears, suggesting
strong interactions between heat carriers. But this is apparently
not the case for $\sigma=2$: the signature of the localized
excitations is still recognized, but probably due to their weak
interactions, their identification now becomes weaker. Both dynamics
in zero- and finite-temperature systems are in good accord with our
conjecture.

\section{Conclusion}
To summarize, we have revealed the intrinsic feature of thermal
transport in a LR interacting system. As such we have shown that a
theoretical model of the LR interacting FPU chain (with Born-von
Karman periodic boundary conditions and under an appropriate range
value $\sigma=2$) can support a higher length-divergent exponent
$\alpha \simeq 0.71$ of the thermal conductivity. This finding is of
fundamental importance as it provides a theoretical possibility to
search for higher thermal conductivity in 1D materials involving LR
interactions, thus pointing towards new manipulations of heat in
practice~\cite{Rect-1,Rect-2}. It also opens up new avenues for
exploring thermal transport as the new divergence surely indicates
new mechanisms.

The higher $\alpha$ is related to the system's more rapid heat
propagation, weaker chaotic dynamics, and also the new dynamics of
traveling DBs (we have also measured the system's equilibrium heat
current auto-correlation function, which supports this higher
$\alpha$ as well [see Appendix B]). A new shaped heat propagating
density is found and its scaling $\gamma \simeq 1.29$ can be well
connected to $\alpha$ by the formula
$\alpha=2-\gamma$~\cite{Report2015}. This seems to indicate that the
L\'{e}vy walk model, of appropriate variations, is still useful for
understanding transport in LR interacting systems. The system's
weaker nonintegrability is confirmed. This suggests that although
the system's chaotic dynamics is not an ingredient sufficient for
the validity of Fourier's law~\cite{Lepri1997}, the strength of
nonintegrability does influence the system's thermal conduction,
thus paving a new way to use nonlinearity to control transport.
More-interestingly, this weaker nonintegrability can result in
peculiar dynamics of traveling DBs at thermal equilibrium and this
seems responsible for the higher $\alpha$. All of these would
undoubtedly encourage further studies of thermal transport.

{\it Final remark:} As our finding ($\alpha \simeq 0.7$) obviously
deviates from those ($\alpha = 1$, ballistic transport)
in~\cite{Bagchi2017-1} by a different protocol of simulations in a
FPU chain with fixed boundary conditions and currently it is
difficult to evaluate whether the results from both protocols are
actually equivalent, one might reconsider this issue in the future.

\begin{acknowledgments}
D.X. would like to acknowledge many helpful discussions with Prof.
R. Livi and Prof. S. Lepri. This work is motivated by their
excellent relevant studies. D.X. is supported by NNSF (Grant No.
11575046) of China and NSF (Grant No. 2017J06002) of Fujian
Province, China. J.W. is supported by NNSF (Grant No. 11847015) of
China and the start-up fund from Jiangxi Science and Technology
Normal University (Grant No. 2017BSQD002). S.V.D. is supported by
the Russian Foundation for Basic Research (Grant No. 19-02-00971).
\end{acknowledgments}

\begin{appendix}
\begin{figure}
\begin{centering}
\vspace{-.6cm}
\includegraphics[width=8.8cm]{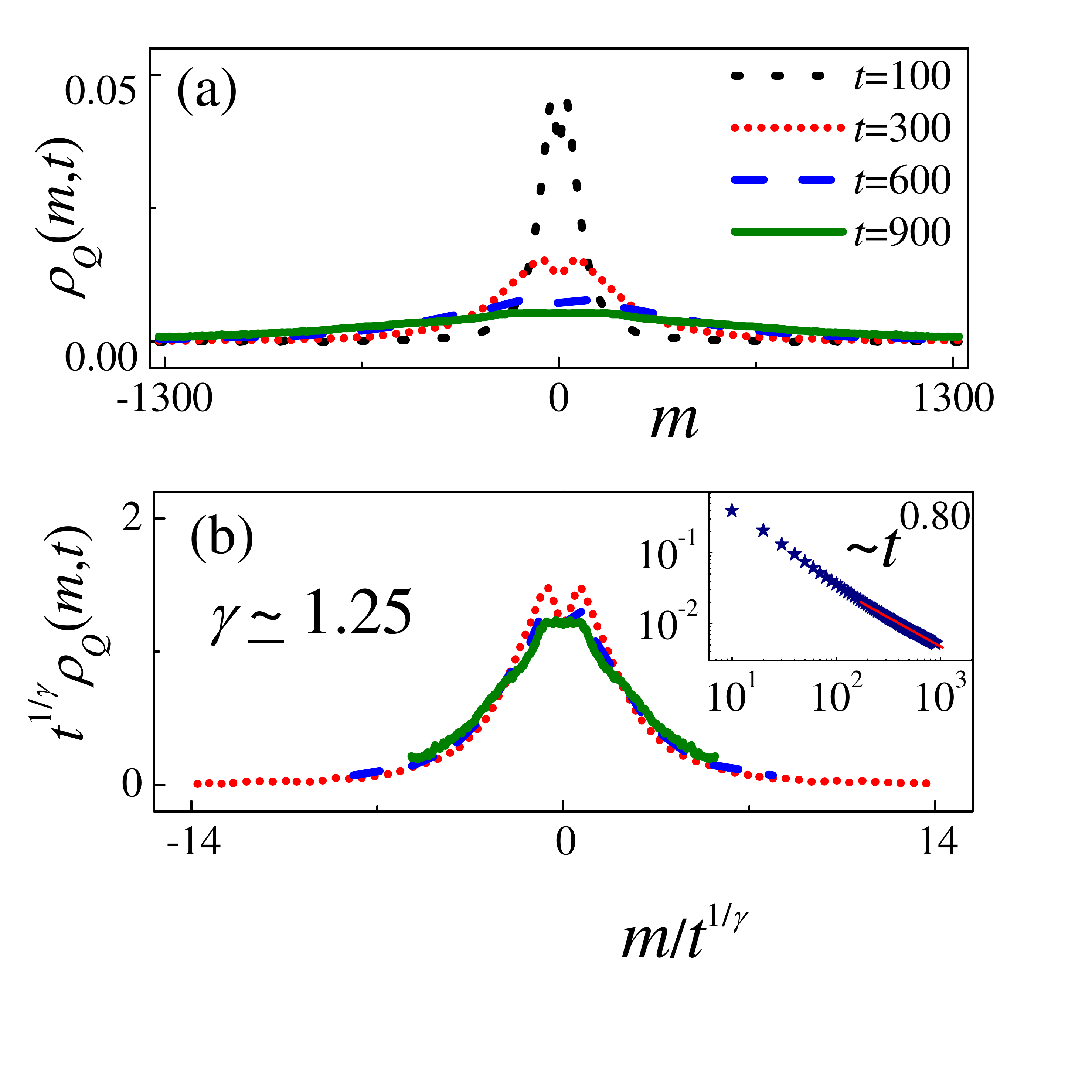} \vspace{-.5cm}
\caption{\label{SFig1} (a) $\rho_Q(m,t)$ for several $t$ for $\sigma=\sigma'=2$. (b) gives the corresponding rescaled $\rho_Q(m,t)$ indicating $\gamma \simeq 1.25$. The inset in (b) shows $\rho_Q(0,t)$ vs t to obtain $\gamma$.}
\end{centering}
\end{figure}
\begin{figure}
\begin{centering}
\vspace{-.4cm} \includegraphics[width=8.8cm]{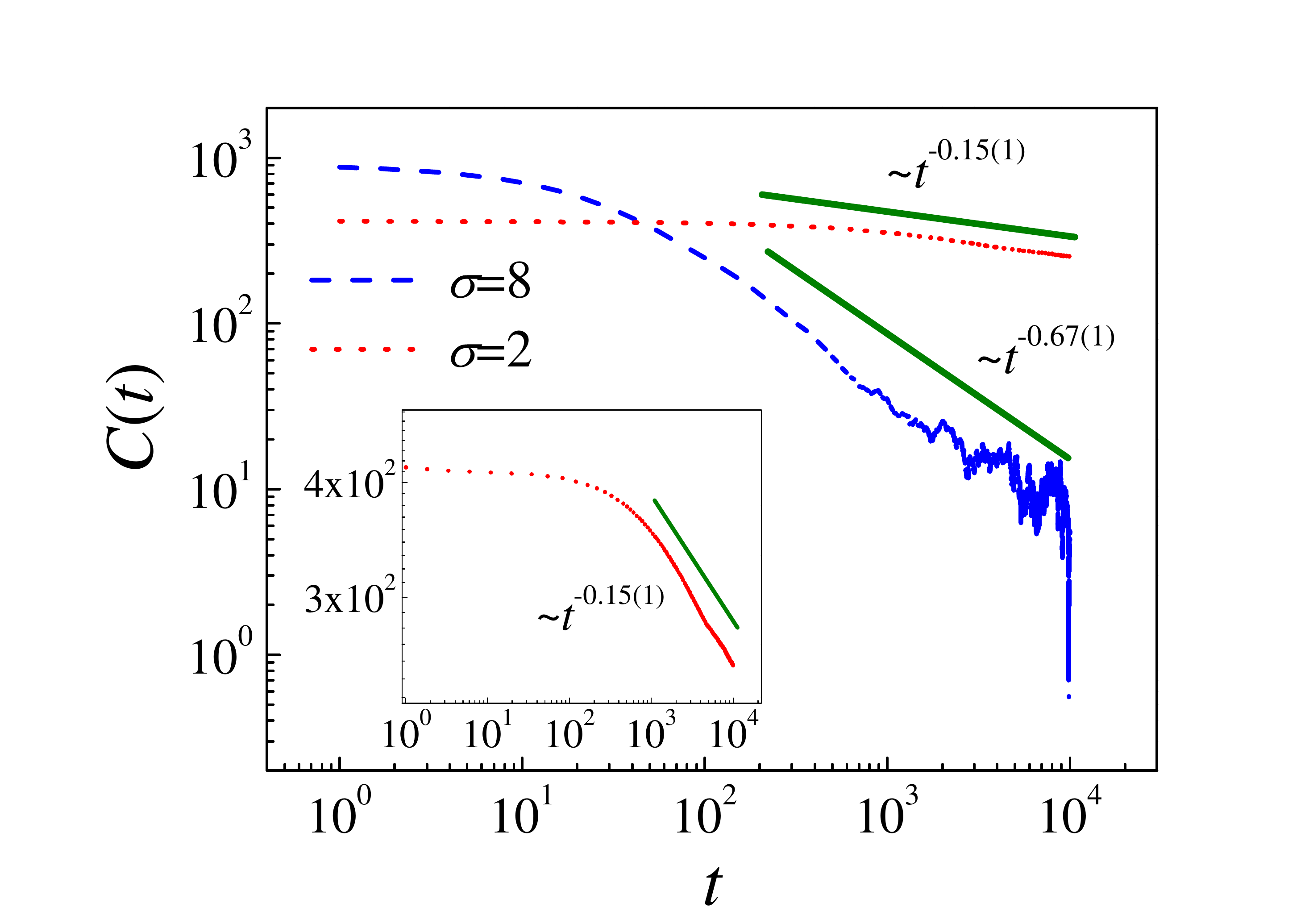} \vspace{-.4cm}
\caption{\label{SFig2} The equilibrium heat current auto-correlation $C(t)$ vs $t$ for $\sigma=2$ and $8$. The inset is a zoom for the result of $\sigma=2$.} \vspace{-.3cm}
\end{centering}
\end{figure}
\section{FPU model with both quadratic and quartic LR interactions}
Including the LR interactions also in the quadratic term, i.e.,
setting $\sigma=\sigma'=2$ in Hamiltonian (1), will not violate our
general conclusion. To demonstrate this, we here present the
relevant results for heat propagation in Fig.~\ref{SFig1}. As can be
seen, $\rho_Q(m,t)$ in Fig.~\ref{SFig1}(a) shows similar shapes to
those in Fig.~\ref{fig:2}(a). A slight difference is that, to
clearly see the platform, longer times are required. A scaling
analysis of $\rho_Q(m,t)$ for different $t$ indicates an exponent
$\gamma \simeq 1.25$ implying $\alpha \simeq 2-1.25=0.75$. This
$\alpha$ value is close to $\alpha \simeq 0.71$ as reported in the
main text.

\section{Equilibrium heat current auto-correlation}
At present, there are three main approaches to detect anomalous
thermal transport behavior, i.e., (i) the direct nonequilibrium
molecular dynamics simulations to obtain $\kappa (L)$; (ii) the
perturbation correlation method to derive $\rho_Q(m,t)$; (iii) the
study of the system's equilibrium heat current auto-correlation
function $C(t)$.

So far we have already shown the results of $\kappa (L)$ and
$\rho_Q(m,t)$ in Figs.~\ref{fig:1} and \ref{fig:2}. To make our
results convincing, here we provide the estimation of $C(t)$. $C(t)$
is defined by
\begin{equation}
C(t)=\langle J_{\rm tot} (t) J_{\rm tot} \rangle,
\end{equation}
where $J_{\rm tot}$ is the total heat current and $\langle \cdot \rangle$ represents the equilibrium average. In the system with LR interactions like Hamiltonian (1) of $\sigma'=\infty$ and $\sigma=2$,
\begin{equation}
J_{\rm tot}=\sum_i p_i \left[x_{i+1}-x_i+ \sum_{j > i} \frac{(x_j-x_i)^3}{(r_{ij})^\sigma}\right].
\end{equation}
The anomalous thermal transport then is related to the slow time decay of $C(t)$:
\begin{equation}
C(t) \sim t^{-\beta}, \qquad  0<\beta<1.
\end{equation}
Using the Green-Kubo formula
\begin{equation}
\kappa=\lim_{\tau \rightarrow \infty} \lim_{N \rightarrow \infty} \frac{1}{k_B N T^2} \int_{0}^{\tau} C(t) dt,
\end{equation}
this slow decay leads to the diverging thermal conductivity $\kappa$.

Figure~\ref{SFig2} depicts the results of $C(t)$ vs $t$ for $\sigma=2$ and $8$. For $\sigma=8$, it indicates $\beta \simeq 0.67$, within the recent one predicted universality class of $\beta=2/3$~\cite{Spohn2014}. In contrast, in the case of $\sigma=2$, a slower decay ($\beta \simeq 0.15$) can be clearly identified. This suggests a new exponent of $\beta$, again supporting our findings and proposed mechanisms.
\end{appendix}


\begin{thebibliography}{100}
\bibitem{Note}
Our study of the model including LR interactions in both the quadratic and quartic terms confirms this, see Appendix A for the model's heat propagation.
\bibitem{Toda1} To facilitate the simulations, we use such a Hamiltonian: $H=\sum_{i} \frac{p_{i}^{2}}{2}+ \exp(x_i-x_{i+1})+x_{i+1}-x_i-1$.
\end{thebibliography}
\end{document}